\documentstyle[e-e-ijmpa,citepunct,twoside,epsfig]{article}

\renewcommand{\thefootnote}{\fnsymbol{footnote}}        %USE SYMBOLIC FOOTNOTE
\def\vev#1{\left\langle #1\right\rangle}

\begin{document}

\normalsize\textlineskip
\pagestyle{empty}

\title{\bf SIMULATION OF LASER-COMPTON COOLING \\
           OF ELECTRON BEAMS%
\footnote{This work was supported in part by the U.S. Department of
Energy under Contract No.~DE-AC03-76SF00098.}
}

\author{TOMOMI OHGAKI}

\address{Lawrence Berkeley National Laboratory \\
         Berkeley, California 94720, USA}

\maketitle\abstracts{We study a method of laser-Compton cooling of
electron beams. Using a Monte Carlo code, we evaluate the effects of
the laser-electron interaction for transverse cooling. The optics with
and without chromatic correction for the cooling are examined. The
laser-Compton cooling for JLC/NLC at $E_0=2$ GeV is considered.}

\setcounter{footnote}{0}
\renewcommand{\thefootnote}{\alph{footnote}}

%--------------------------------------------------------------------

\vspace*{20pt}\textlineskip      %) USE THIS MEASUREMENT WHEN THERE IS
\section{Introduction}          %) A SECTION HEADING
\vspace*{-0.5pt}

  A novel method of electron beam cooling for future linear colliders
was proposed by V.Telnov~\cite{tel96}. During head-on collisions with
laser photons, the transverse distribution of electron beams remains
almost unchanged and also the angular spread is almost
constant. Because the Compton scattered photons follow the initial
electron trajectory with a small additional spread due to much lower
energy of photons (a few eV) than the energy of electrons (several
GeV). The emittance $\epsilon_i=\sigma_i\sigma_i'$ remains almost
unchanged ($i=x,y$). At the same time, the electron energy decreases
from $E_0$ to $E_f$. Thus the normalized emittances have decreased as
follows
\vspace*{5pt}
\begin{equation}
\label{eq:1001}
  \epsilon_n=\gamma\epsilon=\epsilon_{n0}(E_f/E_0)=\epsilon_{n0}/C,
\vspace*{5pt}
\end{equation}
where $\epsilon_{n0}$, $\epsilon_n$ are the initial and final
normalized emittances, the factor of the emittance reduction
$C=E_0/E_f$. The method of electron beam cooling allows further
reduction of the transverse emittances after damping rings or guns by
1-3 orders of magnitude~\cite{tel96}.

  In this paper, we have evaluated the effects of the laser-Compton
interaction for transverse cooling using the Monte Carlo
code CAIN~\cite{che97}. The simulation calculates the effects of the
nonlinear Compton scattering between the laser photons and the
electrons during a multi-cooling stage. Next, we examine the optics
for cooling with and without chromatic correction. The laser-Compton
cooling for JLC/NLC~\cite{ilc96} at $E_0=2$ GeV is considered in
section~4. A summary of conclusion is given in section~5.

%\pagebreak

\textheight=7.8truein
\setcounter{footnote}{0}
\renewcommand{\thefootnote}{\alph{footnote}}
\newpage
%--------------------------------------------------------------------
%   Table 1 : Electron beam parameters
%--------------------------------------------------------------------
\begin{table}[htbp]
\tcaption{Parameters of the electron beams for laser-Compton
cooling. The value in the parentheses is given by Telnov's formulas.}
\centerline{\footnotesize\smalllineskip
\label{tbl:peb}
\begin{tabular}{c c c c c c c}\\
\hline
 $E_0$ (GeV) & $E_f$ (GeV) & $C$ & $\epsilon_{n,x}/\epsilon_{n,y}$
(m$\cdot$\rm{rad}) & $\beta_x/\beta_y$ (mm) & $\sigma_z$ (mm) & $\delta$ (\%) \\
\hline
 2 & 0.2 & 10 & $7.4\times10^{-8}/2.9\times10^{-8}$ &     4/4 & 0.5 &
 11 (9.8) \\
 5 &   1 &  5 & $3.0\times10^{-6}/3.0\times10^{-6}$ & 0.1/0.1 & 0.2 &
 19 (19) \\
\hline\\
\end{tabular}}
\end{table}
%--------------------------------------------------------------------
\vspace*{-20pt}
%--------------------------------------------------------------------
%   Table 2 : Laser beam parameters
%--------------------------------------------------------------------
\begin{table}[htbp]
\tcaption{Parameters of the laser beams for laser-Compton cooling. The 
value in the parentheses is given by Telnov's formulas.}
\centerline{\footnotesize\smalllineskip
\label{tbl:plb}
\begin{tabular}{c c c c c c c}\\
\hline
 $E_0$ (GeV) & $\lambda_L$ ($\mu$m) & $x_0$ & $A$ (J) & $\xi$ & $R_{L,x}/R_{L,y}$ (mm) & $\sigma_{L,z}$ (mm) \\
\hline
 2 & 0.5 & 0.076 & 300 (56) & 2.1 (2.2) & 0.3/0.3 & 1.25 \\
 5 & 0.5 &  0.19 &   20 (4) & 1.5 (1.5) & 0.1/0.1 &  0.4 \\
\hline\\
\end{tabular}}
\end{table}
%--------------------------------------------------------------------
\vspace*{-20pt}
\section{Laser-Electron Interaction}            %) A SECTION HEADING

\subsection{Laser-Electron Interaction}

  In this section, we describe the main parameters for laser-Compton
cooling of electron beams. A laser photon of energy $\omega_L$
(wavelength $\lambda_L$) is backward-Compton scattered by an electron
beam of energy $E_0$ in the interaction point (IP). The kinematics of
Compton scattering is characterized by the dimensionless
parameter~\cite{tel96}
\begin{equation}
\label{eq:2101}
  x_0\equiv\frac{4E_0\omega_L}{m_e^2c^4}=0.019\frac{E_0[\rm
  GeV]}{\lambda_L[\mu\rm m]},
\end{equation}
where $m_e$ is electron mass. The parameters of the electron and laser
beams for laser-Compton cooling are listed in Tables~\ref{tbl:peb}
and~\ref{tbl:plb}. The parameters of the electron beam with 2 GeV are
given for JLC/NLC case in section~4. The parameters of that with 5 GeV
are used for simulation in the next subsection. The wavelength of
laser is assumed to be 0.5 $\mu$m. The parameters of $x_0$ with the
electron energies 2 GeV and 5 GeV are 0.076 and 0.19,
respectively. 

  The required laser flush energy with $Z_R\ll l_{\gamma}\simeq l_e$
is~\cite{tel96}
\begin{equation}
\label{eq:2102}
  A=25\frac{l_e[\rm mm]\lambda_L[\mu\rm m]}{E_0[\rm GeV]}(C-1)~[\rm J],
\end{equation}
where $Z_R$, $l_{\gamma} (\sim 2\sigma_{L,z})$, and $l_e (\sim
2\sigma_z)$ are the Rayleigh length of laser, and the bunch lengths of
laser and electron beams. From this formula, the parameters of $A$
with the electron energies 2 GeV and 5 GeV are 56 J and 4 J,
respectively. 

  The nonlinear parameter of laser field is~\cite{tel96}
\begin{equation}
\label{eq:2103}
  \xi^2=4.3\frac{\lambda_L^2[\mu\rm m^2]}{l_e [\rm mm]\it E_{\rm0}[\rm
  GeV]}(C-1).
\end{equation}
In this study, for the electron energies 2 GeV and 5 GeV, the
parameters of $\xi$ are 2.2 and 1.5, respectively. 

  The rms energy of the electron beam after Compton scattering
is~\cite{tel96}
\begin{equation}
\label{eq:2104}
  \sigma_e=\frac1{C^2}\left[\sigma_{e0}^2[\rm GeV^2]+0.7\it{x}_{\rm0
  }(\rm1+0.45\xi)(\it C-\rm1)\it E_{\rm 0}^{\rm2}[\rm
  GeV^2]\right]^{1/2}~[\rm GeV],
\end{equation}
where the rms energy of the initial beam is $\sigma_{e0}$ and the
ratio of energy spread is defined as $\delta=\sigma_e/E_f$. If the
parameter $\xi$ or $x_0$ is larger, the energy spread after Compton
scattering is increasing and it is the origin of the emittance growth
in the defocusing optics, reacceleration linac, and focusing
optics. The energy spreads $\delta$ for the electron energies 2 GeV
and 5 GeV are 9.8\% and 19\%, respectively.

  The equilibrium emittances due to Compton scattering are~\cite{tel96}
\begin{equation}
\label{eq:2105}
  \epsilon_{ni,\min}=\frac{7.2\times10^{-10}\beta_i[\rm
  mm]}{\lambda_L[\mu\rm m]}~(\it i=x,y)~[\rm m\cdot rad],
\end{equation}
where $\beta_i$ are the beta functions at IP. From this formula we can
see that small beta gives small emittance. However the large change of
the beta functions between the magnet and the IP causes the emittance
growth. Taking no account of the emittance growth, for the electron
energies 2 GeV and 5 GeV, the equilibrium emittances are
$5.8\times10^{-9}$ m$\cdot$rad and $1.4\times10^{-10}$ m$\cdot$rad,
respectively. The equilibrium emittances depended on $\xi$ in the case
$\xi^2\gg1$ were calculated in Ref.~\citenum{tel96}.

\subsection{Simulation of Laser-Electron Interaction}

  For the simulation of laser-electron interaction, the electron beam
is simply assumed to be a round beam in the case of $E_0=5$ GeV and
$C=5$. Taking no account of the emittance growth of optics, the one
stage for cooling consists two parts as follows:
\begin{enumerate}
  \item The laser-Compton interaction between the electron and laser
  beams.
  \item The reacceleration of electrons in linac.
\end{enumerate}
In the first part, we simulated the interactions by the CAIN
code~\cite{che97}. This simulation calculates the effects of the
nonlinear Compton scattering between the laser photons and the
electrons. We assume that the laser pulse interacts with the electron
bunch in head-on collisions. The $\beta_x$ and $\beta_y$ at the IP
are fixed to be 0.1 mm. The initial energy spread of the electron
beams is 1\%. The energy of laser pulse is 20~J. The difference of the
pulse energy between the simulation and the formula depends on the
transverse sizes of the electron beams at IP. The polarization of the
electron and laser beams are $P_e$=1.0 and $P_L$=1.0 (circular
polarization), respectively. When the $x_0$ and $\xi$ parameters are
small, the spectrum of the scattered photons does not largely depend
on the polarization combination. In order to accelerate the electron
beams to 5 GeV for recovery of energy in the second part, we simply
added the energy $\Delta E=5~\mbox{GeV}-E_{ave}$ for reacceleration,
where $E_{ave}$ is the average energy of the scattered electron beams
after the laser-Compton interaction.

  Here we define the transverse, longitudinal, and 6D emittances in
the simulation. The $x,y$-transverse emittance is
\begin{equation}
\label{eq:2201}
  \epsilon_{n,i}=\sqrt{\sigma_i^2 \sigma_{i'}^2-(\vev{i\cdot
  i'}-\vev{i}\vev{i'})^2} \ \ (\it i=x,y),
\end{equation}
where the symbol $\vev{ \ }$ means to take an average of all particles
in a bunch. 

  The longitudinal emittance is
\begin{equation}
\label{eq:2203}
  \epsilon_{n,l}=\sqrt{\sigma_z^2 \sigma_{p_z}^2-(\vev{z\cdot
  p_z}-\vev{z}\vev{p_z})^2}/(m_e c).
\end{equation}

  The 6D emittance is defined as
\begin{equation}
\label{eq:2204}
  \epsilon_{6N}=\epsilon_{n,x}\cdot\epsilon_{n,y}\cdot\epsilon_{n,l}.
\end{equation}

\begin{figure}[htbp]
\vspace*{0pt}
\centerline{\epsfig{file=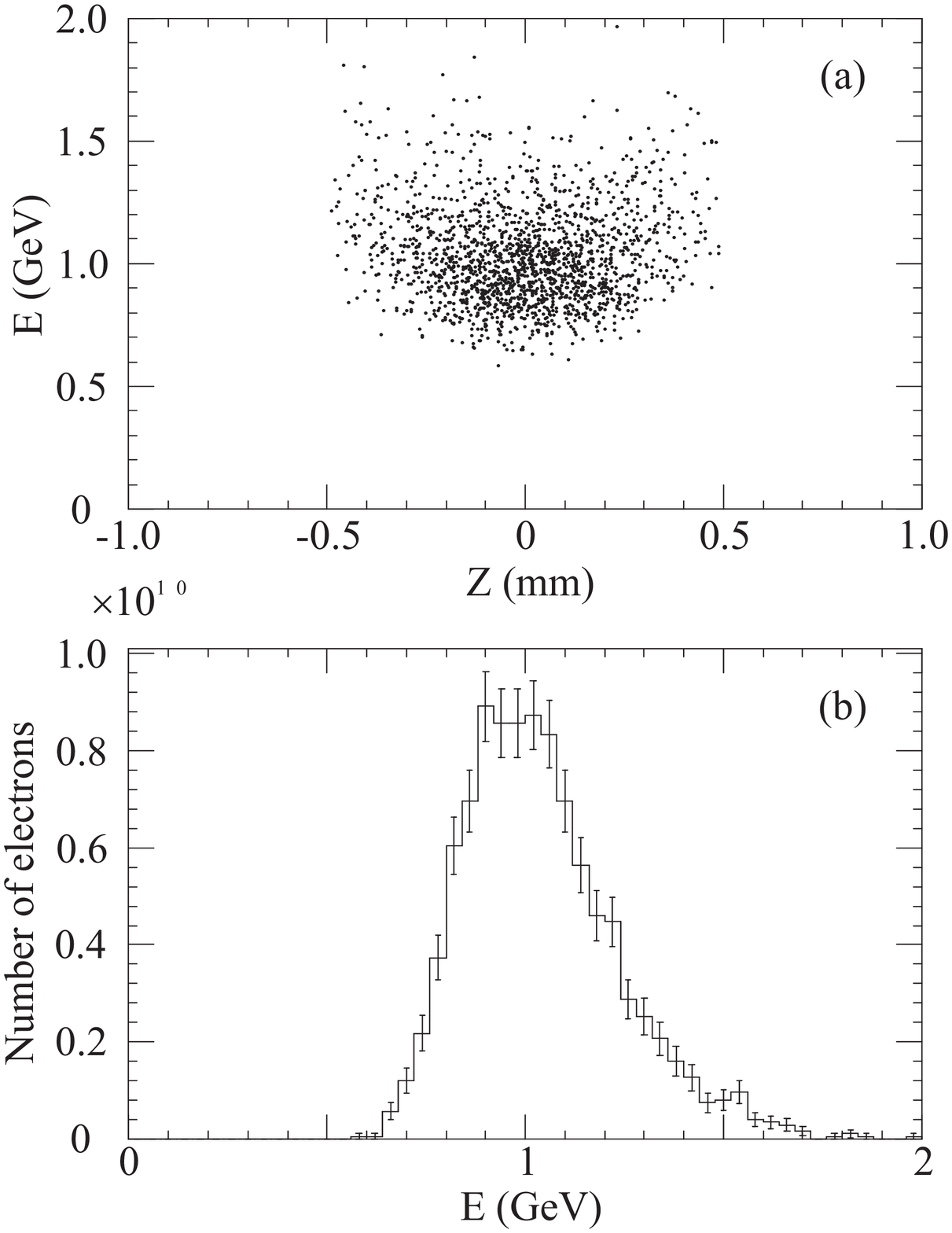,width=4.5cm}}
\vspace*{2pt}
\fcaption{The longitudinal distribution of the electrons. (a) The
energy \it{vs.}~$z.$ \rm (b) The energy distribution of the
electrons. The bin size is 40 MeV.}
\label{fig:lde}
\end{figure}

  Figure~\ref{fig:lde} shows the longitudinal distribution of the
electrons after the first laser-Compton scattering. The average energy
of the electron beams is 1.0 GeV and the energy spread $\delta$ is
0.19. The longitudinal distribution seems to be a boomerang. If we
assume a short Rayleigh length of laser pulse, the energy loss of head
and tail of beams is small. The number of the scattered photons per
incoming particle and the average of the photon energy at the first
stage are 40 and 96 MeV (rms energy 140 MeV), respectively.
%--------------------------------------------------------------------
%   Figure 1 : Longitudinal distribution of electrons
%--------------------------------------------------------------------
%--------------------------------------------------------------------

%--------------------------------------------------------------------
%   Figure 2 : Transverse sizes of electron beams
%--------------------------------------------------------------------
\begin{figure}[htbp]
\begin{minipage}[t]{6.3cm}
\vspace*{-4pt}
\centerline{\epsfig{file=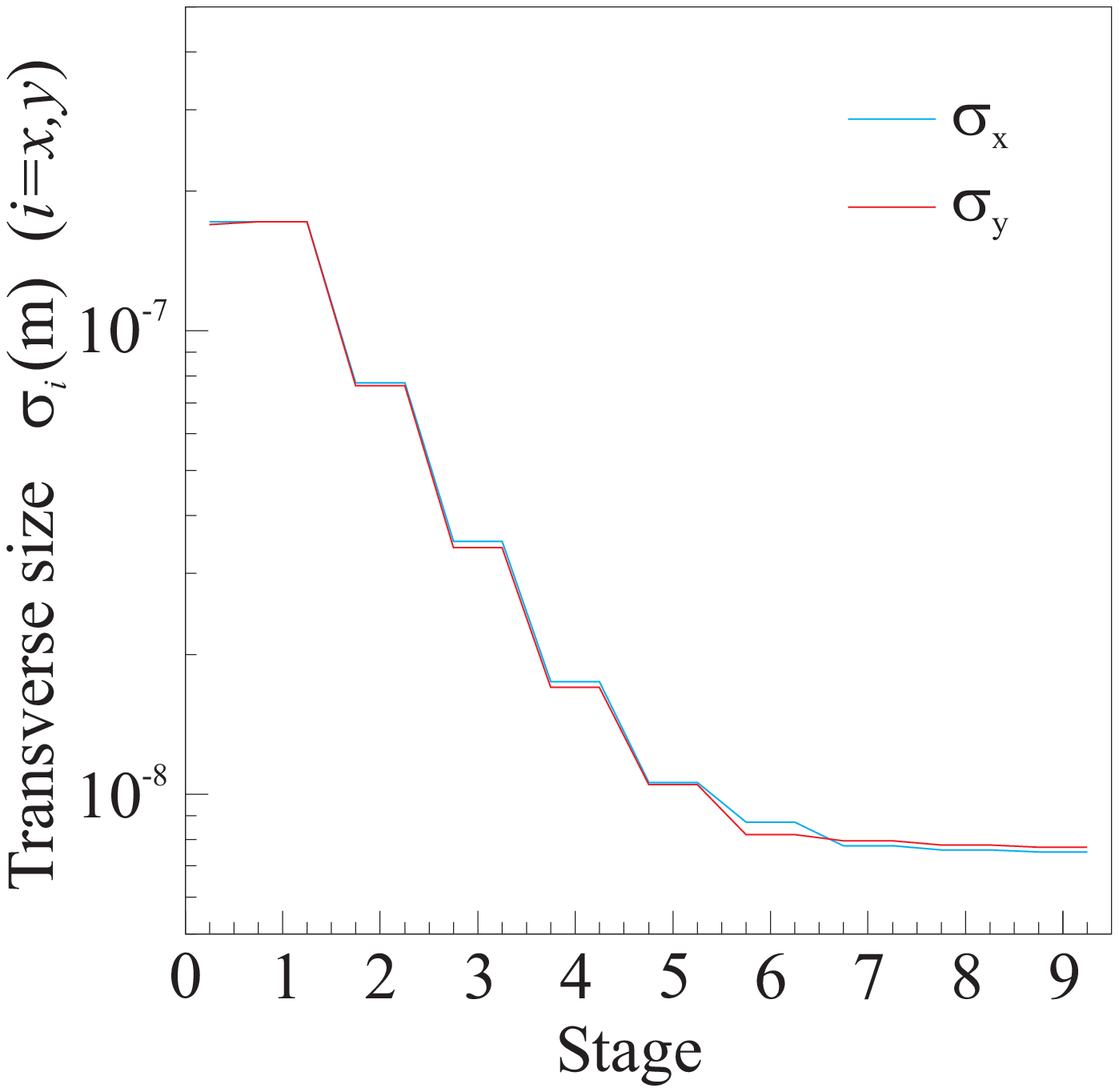,width=5.0cm}}
\vspace*{-4pt}
\fcaption{The transverse sizes of the electron beams.}
\label{fig:tbs}
\end{minipage}
%--------------------------------------------------------------------
%   Figure 3 : Angles of electron beams
%--------------------------------------------------------------------
\begin{minipage}[t]{6.3cm}
\vspace*{-4pt}
\centerline{\epsfig{file=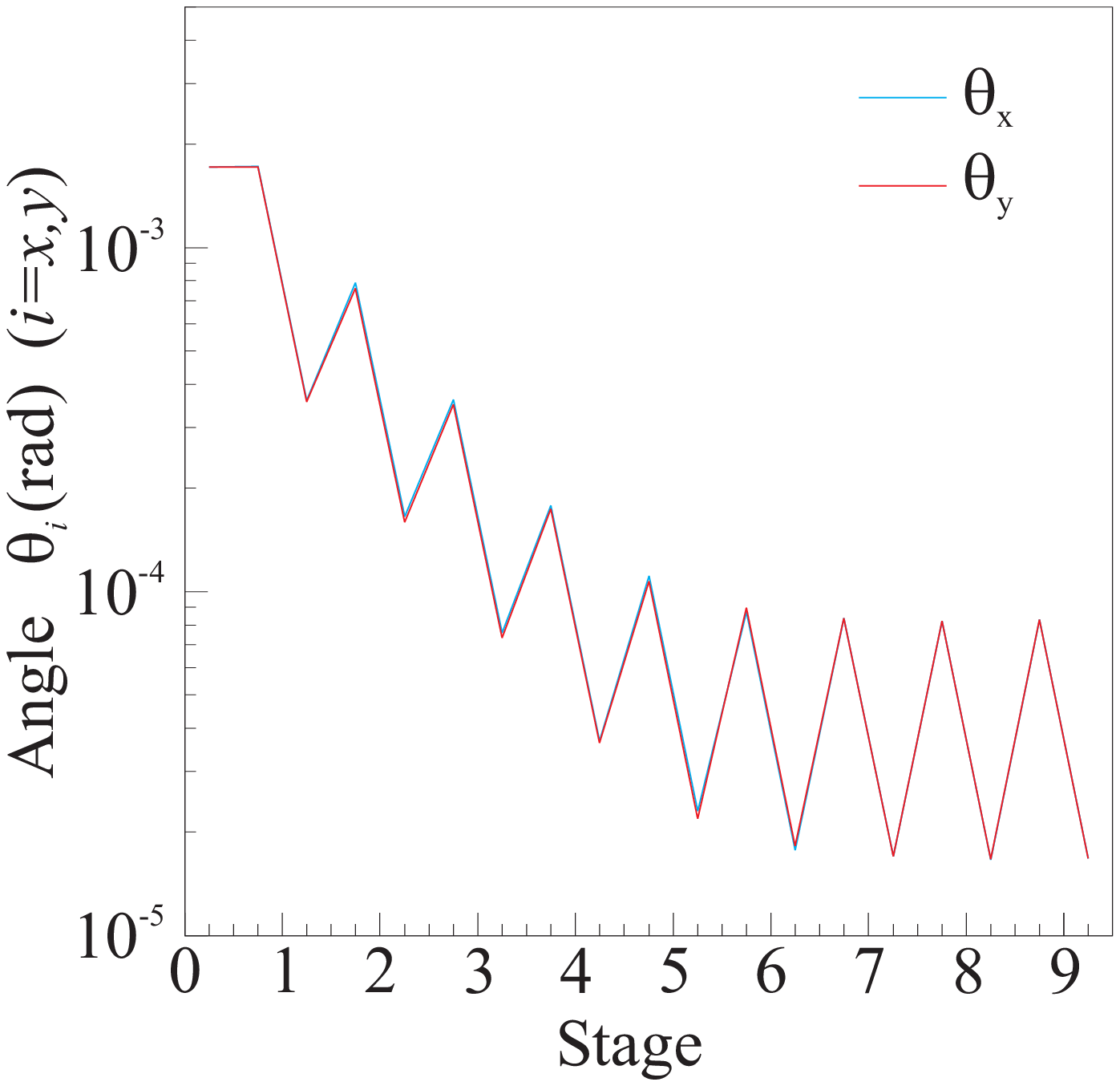,width=5.0cm}}
\vspace*{-4pt}
\fcaption{The angles of the electron beams.}
\label{fig:tab}
\end{minipage}
\vspace*{-10pt}
\end{figure}
%--------------------------------------------------------------------

  The transverse sizes of the electron beams in the multi-stage
cooling are shown in Fig.~\ref{fig:tbs}. During collisions with the
laser photons, the transverse distribution of the electrons remains
almost unchanged. But they decrease when we focus them for the next
laser-Compton interaction due to the lower normalized emittance and
the fixed $\beta$-function at IP
($\sigma_i=\sqrt{\beta_i\epsilon_{n,i}/\gamma}$). The angles of the
electron beams in the multi-stage cooling are shown in
Fig.~\ref{fig:tab}. As a result of reacceleration, the angles of the
electrons decrease. They increase when we focus them for the next
laser-Compton interaction. Finally the angles attain the average of
Compton scattering angle and the effect of cooling saturates.
             
\begin{figure}[htbp]
\begin{minipage}[t]{6.3cm}
\vspace*{-3pt}
\centerline{\epsfig{file=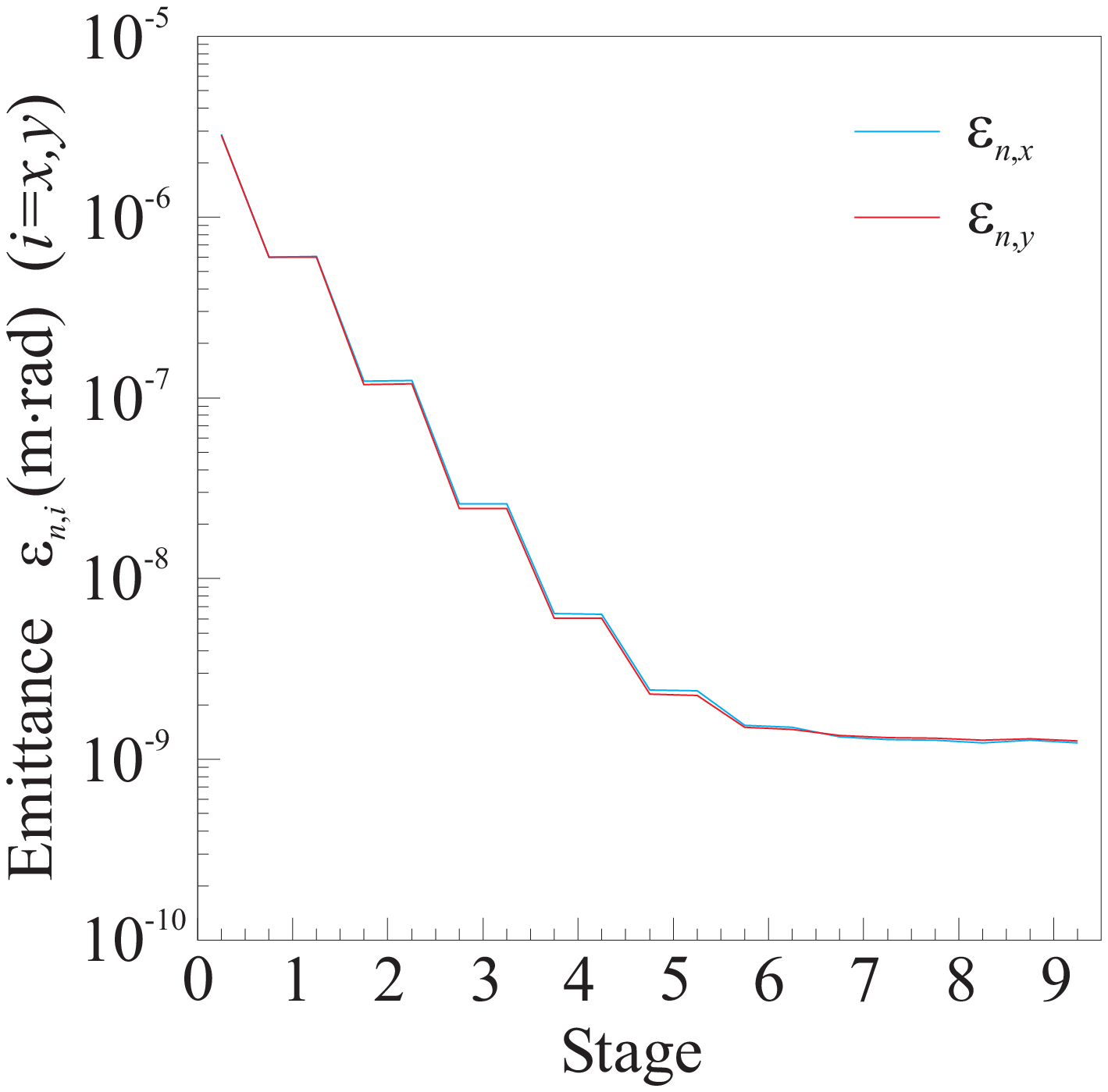,width=5cm}}
\vspace*{-3pt}
\fcaption{The transverse emittances of the electron beams.}
\label{fig:teb}
\end{minipage}
%--------------------------------------------------------------------
%   Figure 5 : Longitudinal emittances of electron beams
%--------------------------------------------------------------------
\begin{minipage}[t]{6.3cm}
\vspace*{-3pt}
\centerline{\epsfig{file=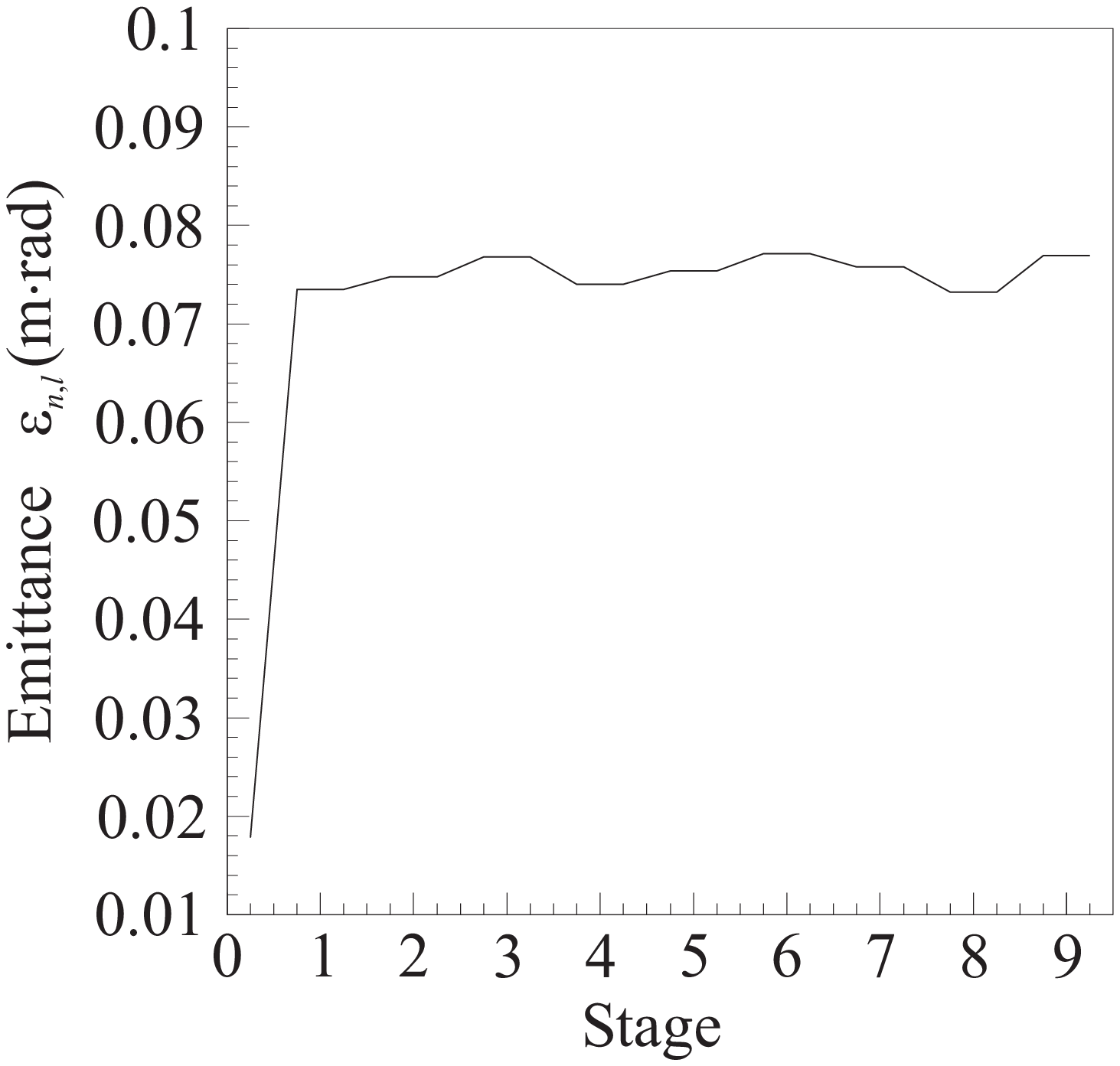,width=5cm}}
\vspace*{-3pt}
\fcaption{The longitudinal emittance of the electron beams.}
\label{fig:leb}
\end{minipage}
\end{figure}

\begin{figure}[htbp]
\begin{minipage}[t]{6.3cm}
\vspace*{-4pt}
\centerline{\epsfig{file=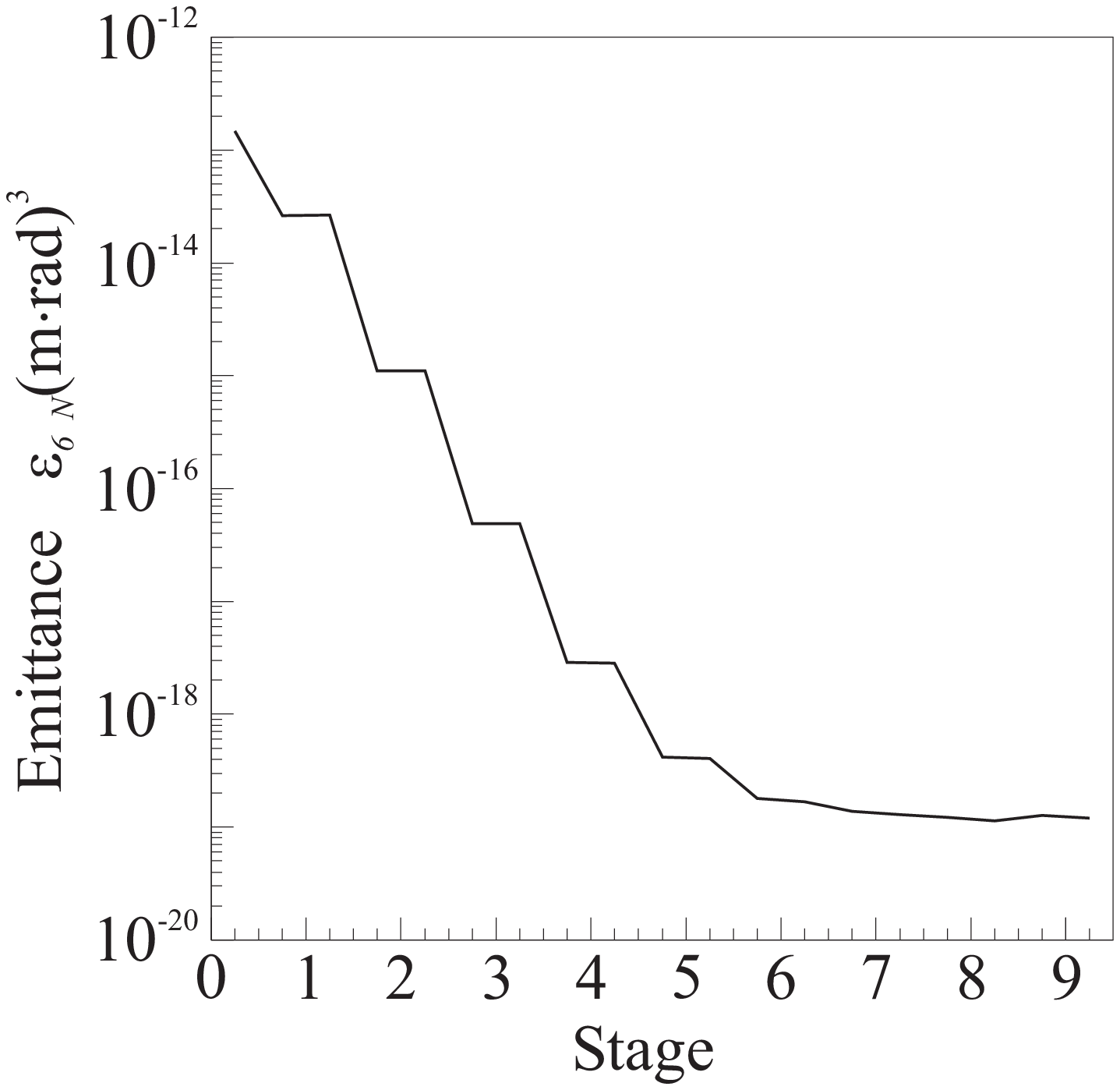,width=5cm}}
\vspace*{-4pt}
\fcaption{The 6D emittance of the electron beams.}
\label{fig:6de}
\end{minipage}
%--------------------------------------------------------------------
%   Figure 7 : Polarization of electron beams
%--------------------------------------------------------------------
\begin{minipage}[t]{6.3cm}
\vspace*{-4pt}
\centerline{\epsfig{file=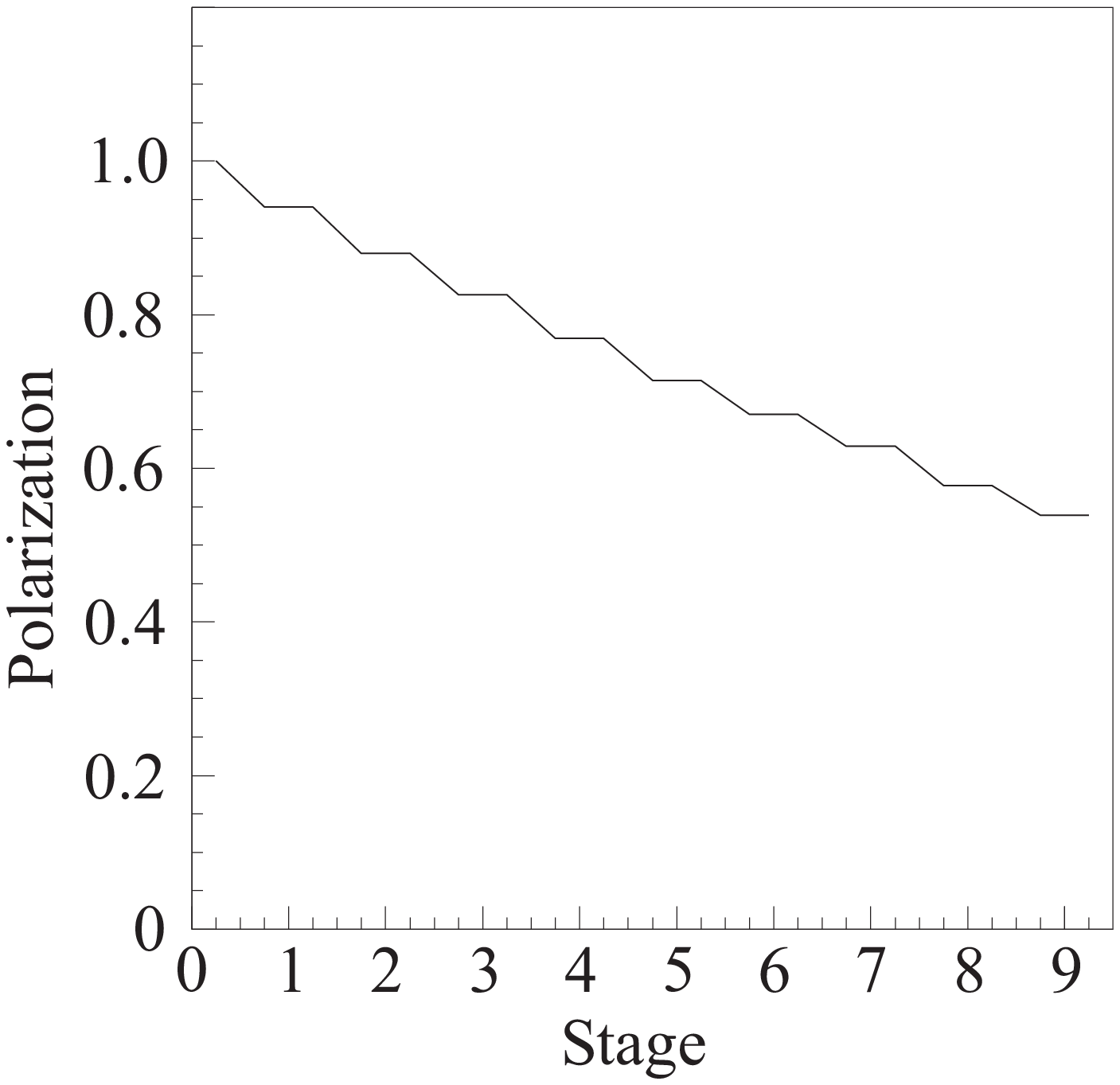,width=5cm}}
\vspace*{-4pt}
\fcaption{The polarization of the electron beams.}
\label{fig:epo}
\end{minipage}
\vspace*{-10pt}
\end{figure}

  Figure~\ref{fig:teb} shows the transverse emittances of the electron
beams in the multi-stage cooling. From Eq.(\ref{eq:2105}),
$\epsilon_{ni,\rm{min}}=1.4\times10^{-10}$ m$\cdot$rad, and the
simulation presents $\epsilon_{ni,\rm{min}}=1.2\times10^{-9}$
m$\cdot$rad. Figure~\ref{fig:leb} shows the longitudinal emittance of
the electron beams in the multi-stage cooling. Due to the increase of
the energy spread of the electron beams from 1\% to 19\%, the
longitudinal emittance rapidly increases at the first stage. After the
first stage, the normalized longitudinal emittance is stable. The 6D
emittance of the electron beams in the multi-stage cooling is shown in
Fig.~\ref{fig:6de}. The second cooling stage has the largest reduction
for cooling. The 8th or 9th cooling stages have small reduction for
cooling. The initial and final 6D emittances $\epsilon_{6N}$ are
$1.5\times10^{-13}$ (m$\cdot$rad)$^3$ and $1.2\times10^{-19}$
(m$\cdot$rad)$^3$, respectively.
             
  Figure~\ref{fig:epo} shows the polarization of the electron beams in
the multi-stage cooling. The decrease of the polarization during the
first stage is about 0.06. The final polarization $P_e$ after the
multi-stage cooling is 0.54.
%--------------------------------------------------------------------
%   Figure 4 : Transverse emittances of electron beams 
%--------------------------------------------------------------------
%--------------------------------------------------------------------
             
%--------------------------------------------------------------------
%   Figure 6 : 6D emittance of electron beams.
%--------------------------------------------------------------------
%--------------------------------------------------------------------

%--------------------------------------------------------------------
\vspace*{-20pt}
\section{Optics Design for Laser-Compton Cooling}               %) A SECTION HEADING

\subsection{Optics without chromaticity correction}

\begin{figure}[htbp]
\vspace*{-3pt}
\centerline{\epsfig{file=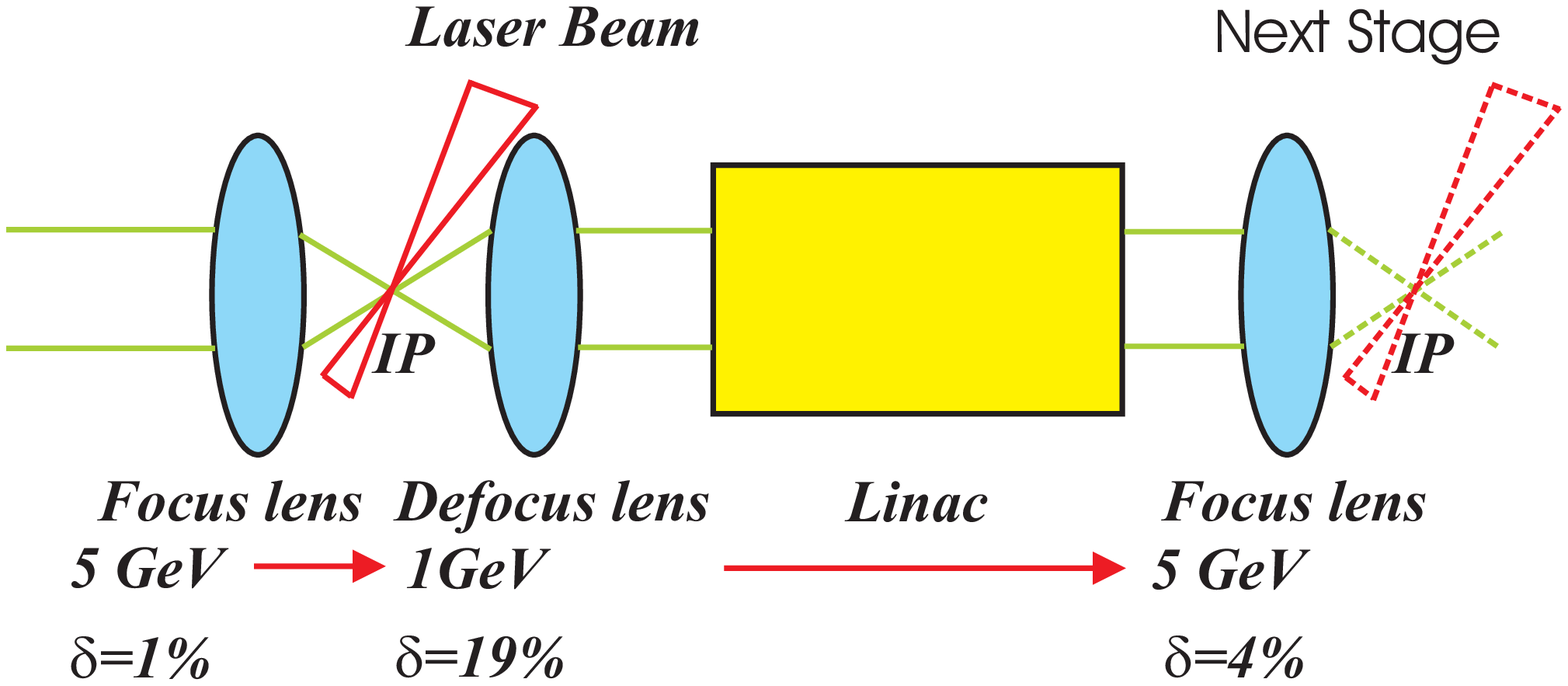,width=6.0cm}}
\fcaption{Schematics of the laser-Compton cooling of electron beams.}
\vspace*{-3pt}
\label{fig:lce}
\end{figure}

  There are three optical devices for the laser-Compton cooling of
electron beams as follows:
\begin{enumerate}
  \item The focus optics to the first IP.
  \item The defocus optics from the first IP to the reacceleration
  linac.
  \item The focus optics from the linac to the next IP.
\end{enumerate}
Figure~\ref{fig:lce} shows schematics of the laser-Compton cooling
of electron beams. The optics 1 is focusing the electron beams from
a few meters of $\beta$-function to several millimeters in order
to effectively interact them with the laser beams. The optics 2 is
defocusing them from several millimeters to a few meters for
reacceleration of electron beams in linac. In a multi-stage cooling
system, the optics 3 is needed for cooling in the next stage. The
problem for the focus and defocus optical devices is the energy spread
of electrons and the electron beams with a large energy spread are
necessary to minimize or correct the chromatic aberrations avoiding
emittance growth.
%--------------------------------------------------------------------
%   Figure 8 : Schematics of the Laser-Compton cooling
%--------------------------------------------------------------------
%--------------------------------------------------------------------

  In this subsection, we discuss the optics for laser-Compton cooling
without chromatic corrections. For the focus and defocus of the beams,
we use the final doublet system which is similar to that of the final
focus system of the future linear colliders~\cite{ilc96}. The pole-tip
field of the final quadrupole $B_T$ is limited to 1.2 T and the
pole-tip radius $a$ is greater than 3 mm. The strength of the final
quadrupole is
\begin{eqnarray}
\label{eq:3101}
  \kappa=B_T/(a B\rho)\le120/E [\rm{GeV}]~[\rm m^{-2}],
\end{eqnarray}
where $B$, $\rho$, and $E$ are the magnetic field, the radius of
curvature and the energy of the electron beams. In our case, the
electron energies in the optics 1, 2, and 3 are 5.0, 1.0, and 5.0 GeV,
respectively and the limit of the strength of the quadrupole in laser
cooling is much larger than that of the final quadrupole of the future
linear colliders. Due to the low energy beams in laser cooling, the
synchrotron radiation from quadrupoles and bends is negligible.

\begin{figure}[htbp]
\begin{minipage}[]{6.3cm}
\vspace*{-3pt}
\centerline{\epsfig{file=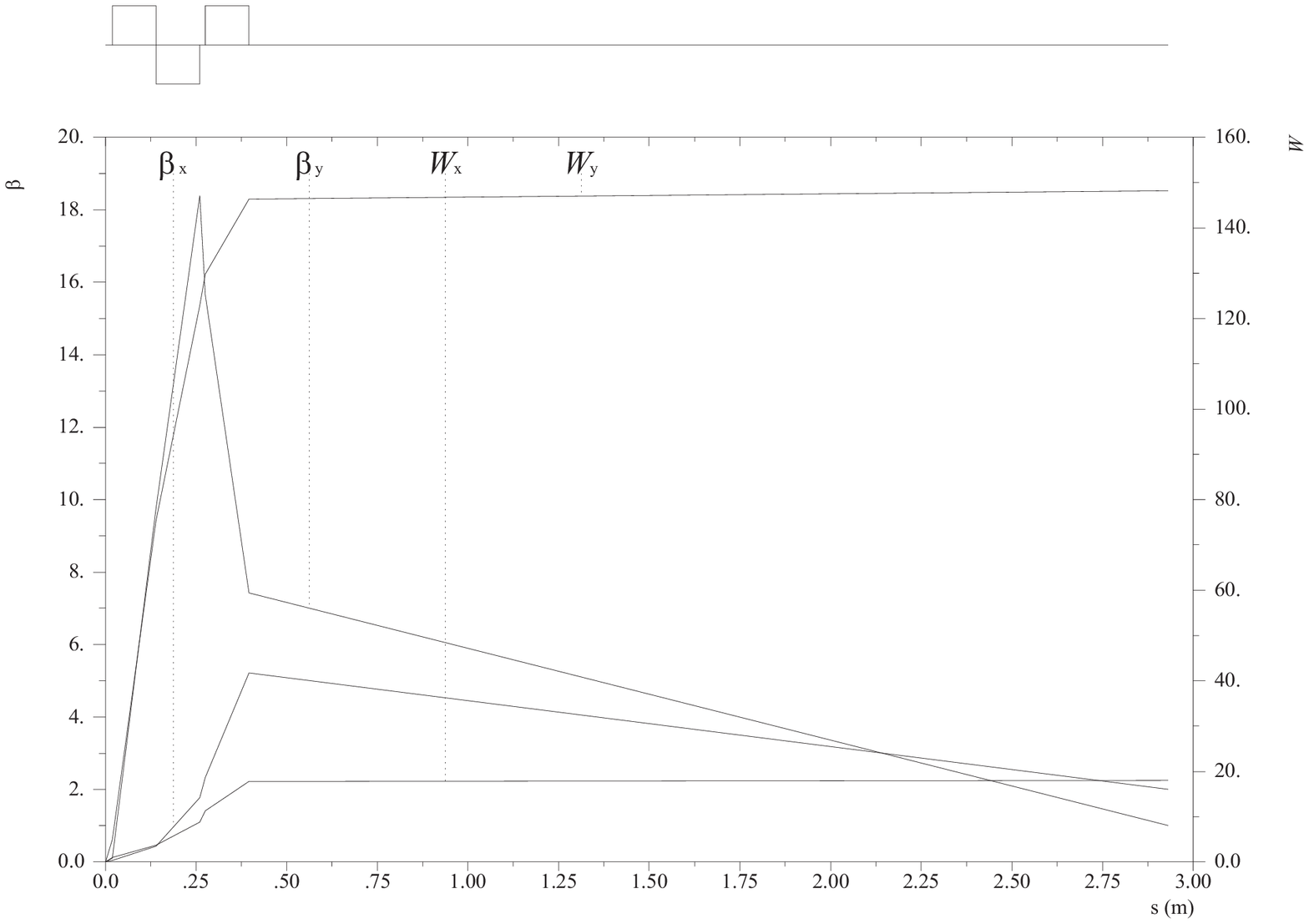,width=6cm}}
\vspace*{-3pt}
\fcaption{The defocus optics without chromaticity correction for
laser-Compton cooling.}
\label{fig:dlo}
\end{minipage}
%--------------------------------------------------------------------
%   Figure 10: Momentum dependence of the emittances in the defocus
%              optics without chromaticity correction
%--------------------------------------------------------------------
\begin{minipage}[]{6.3cm}
\vspace*{-3pt}
\centerline{\epsfig{file=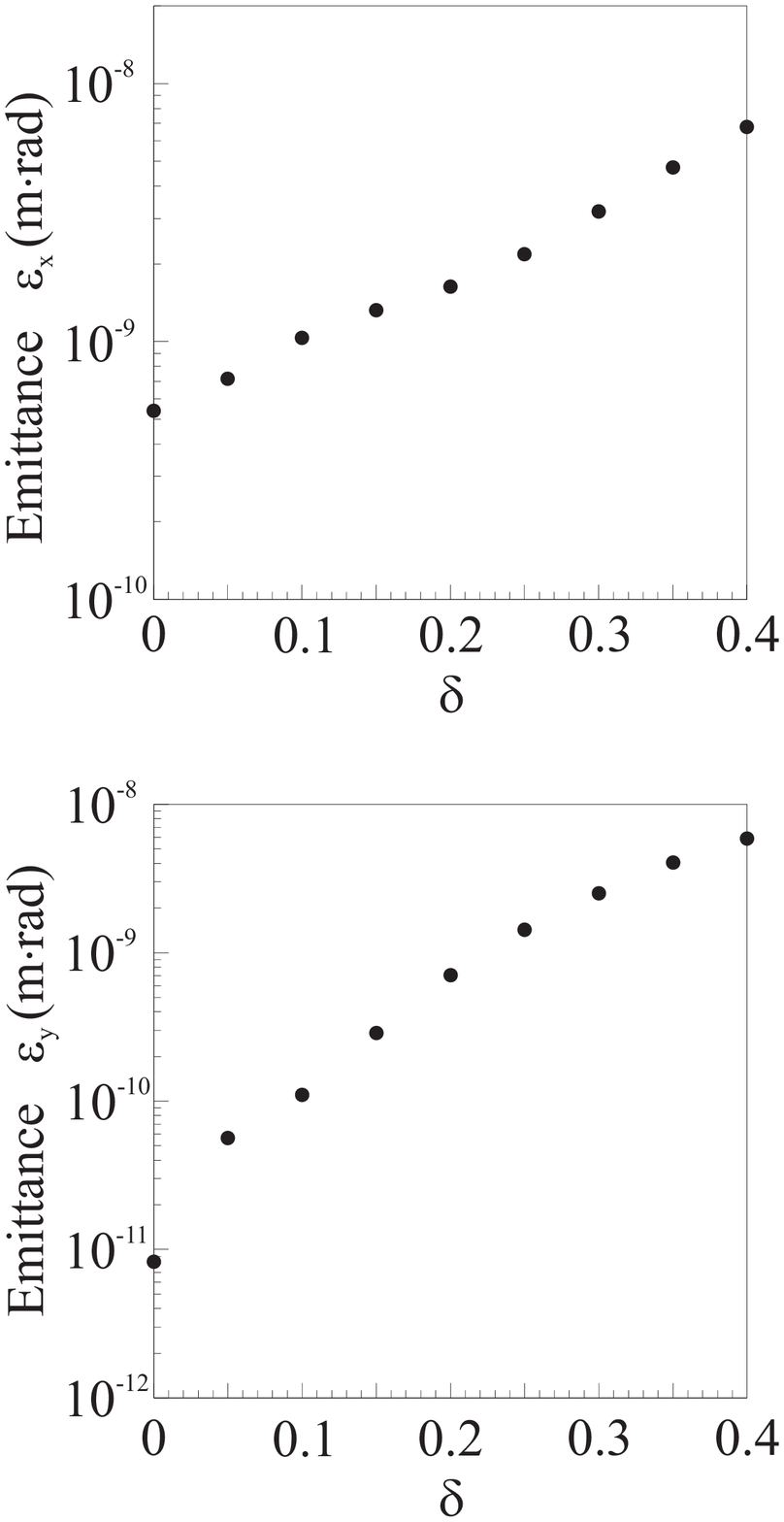,width=4.5cm}}
\vspace*{-3pt}
\fcaption{Momentum dependence of the \\
emittances in the defocus optics without \\
chromaticity correction.}
\label{fig:mdl}
\end{minipage}
\vspace*{-10pt}
\end{figure}

  The difference of three optical devices is the amount of the energy
spread of the beams. In the optics 1,2, and 3, the beams have one,
several tens, and a few~\% energy spread. In order to minimize the
chromatic aberrations, we need to shorten the length between the
final quadrupole and the IP. In this study, the length from the face
of the final quadrupole to the IP, $l$ is assumed to be 2~cm. Here we
estimated the emittance growth in the optics 2, because the chromatic
effect in the optics 2 is the most serious. Figure~\ref{fig:dlo}
shows the defocus optics without chromaticity correction for
laser-Compton cooling by the MAD code~\cite{mad96}. The input file is
attached to Ref.~\citenum{ohg99}. The parameters of the electron beam
for laser-Compton cooling at $E_0=5$ GeV and $C=5$ are listed in
Table~\ref{tbl:peo}. The initial $\beta_x$ and $\beta_y$ after
laser-Compton interaction are 20~mm and 4~mm, respectively. The final
$\beta_x$ and $\beta_y$ are assumed to be 2~m and 1~m,
respectively. The initial and final $\alpha_{x(y)}$ with no energy
spread $\delta=0$ are 0 in this optics. The strength $\kappa$ of the
final quadrupole for the beam energy of 1 GeV from Eq.~(\ref{eq:3101})
is assumed to be 120 $\rm{m}^{-2}$.
%--------------------------------------------------------------------
%   Table 3 : Parameters of electrons for optics design
%--------------------------------------------------------------------
\begin{table}[htbp]
\tcaption{Parameters of the electron beam for laser-Compton cooling
  at $E_0=5$ GeV and $C=5$ for the optics design.}
\centerline{\footnotesize\smalllineskip
\label{tbl:peo}
\begin{tabular}{ccccccccc}\\
\hline
  $E_0$ (GeV) & $\epsilon_{n,x}/\epsilon_{n,y}$ (m$\cdot$rad) & $\beta_x/\beta_y$ (mm) & $\sigma_x/\sigma_y$ (m) & $\sigma_z$ (mm) \\
\hline       
  5 &  $1.06\times 10^{-6}/1.6\times10^{-8}$ & 20/4 &
  $3.3\times10^{-5}/1.8\times10^{-7}$ & 0.2 \\
\hline\\
\end{tabular}}
\vspace*{-5pt}
\end{table}  
%--------------------------------------------------------------------

  In our case, the chromatic functions $\xi_x$ and $\xi_y$ are 18 and
148, respectively. The momentum dependence of the emittances in
the defocus optics without chromaticity correction is shown in
Fig.~\ref{fig:mdl}. In the paper~\cite{mon87}, the analytical study by
thin-lens approximation has been studied for the focusing system, and
here the transverse emittances are calculated by a particle-tracking
simulation. The 10000 particles are tracked for the transverse and
longitudinal Gaussian distribution by the MAD code. The relative
energy spread $\delta$ is changed from 0 to 0.4. Due to the larger
chromaticity $\xi_y$, the emittance $\epsilon_y$ is rapidly increasing
with the energy spread $\delta$. If we set a limit of 200\% for
$\Delta\epsilon_i/\epsilon_i \ (i=x,y)$, the permissible energy spread
$\delta_x$ and $\delta_y$ are 0.11 and 0.012 which mean the momentum
band widths $\pm 22\%$ and $\pm 2.4\%$, respectively. The results are
not sufficient for cooling at $E_0=5$ GeV and $C=5$, because the beams
through the defocusing optics have the energy spread of several
tens~\%. On the one hand, the optics can be useful as the optics 1 and
3 with the energy spread of a few~\%.
%--------------------------------------------------------------------
%   Figure 9 : The defocus optics without chromaticity correction 
%              for the Laser-Compton cooling.
%--------------------------------------------------------------------
%--------------------------------------------------------------------

\subsection{Optics with chromaticity correction}

\begin{figure}[htbp]
\begin{minipage}[]{6.3cm}
\vspace*{-3pt}
\centerline{\epsfig{file=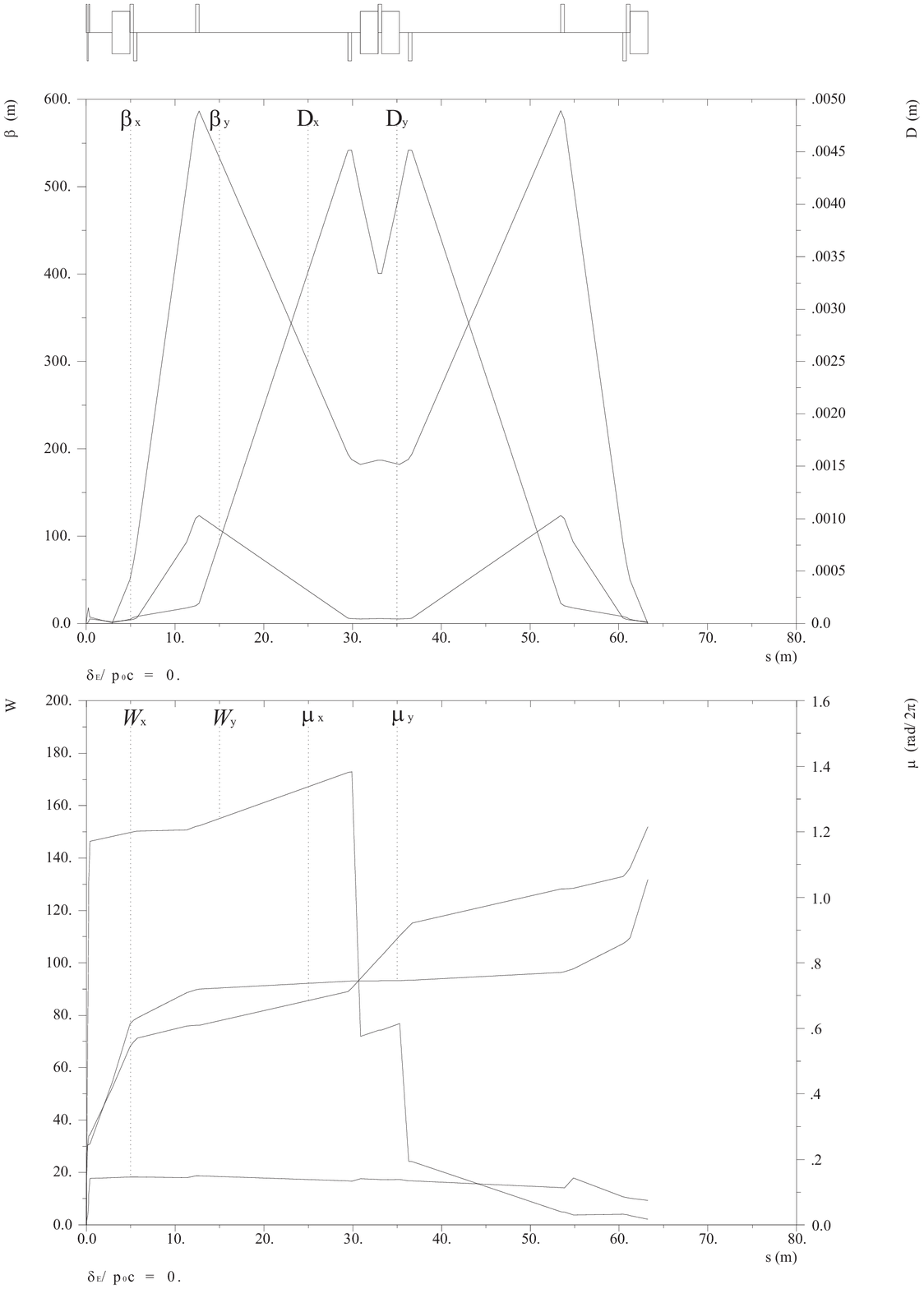,width=6cm}}
\vspace*{-3pt}
\fcaption{The defocus optics without chromaticity correction for
laser-Compton cooling.}
\label{fig:dno}
\end{minipage} 
%--------------------------------------------------------------------
%   Figure 12: Momentum dependence of the emittances in the defocus
%              optics with chromaticity correction
%--------------------------------------------------------------------
\begin{minipage}[]{6.3cm}
\vspace*{-3pt}
\centerline{\epsfig{file=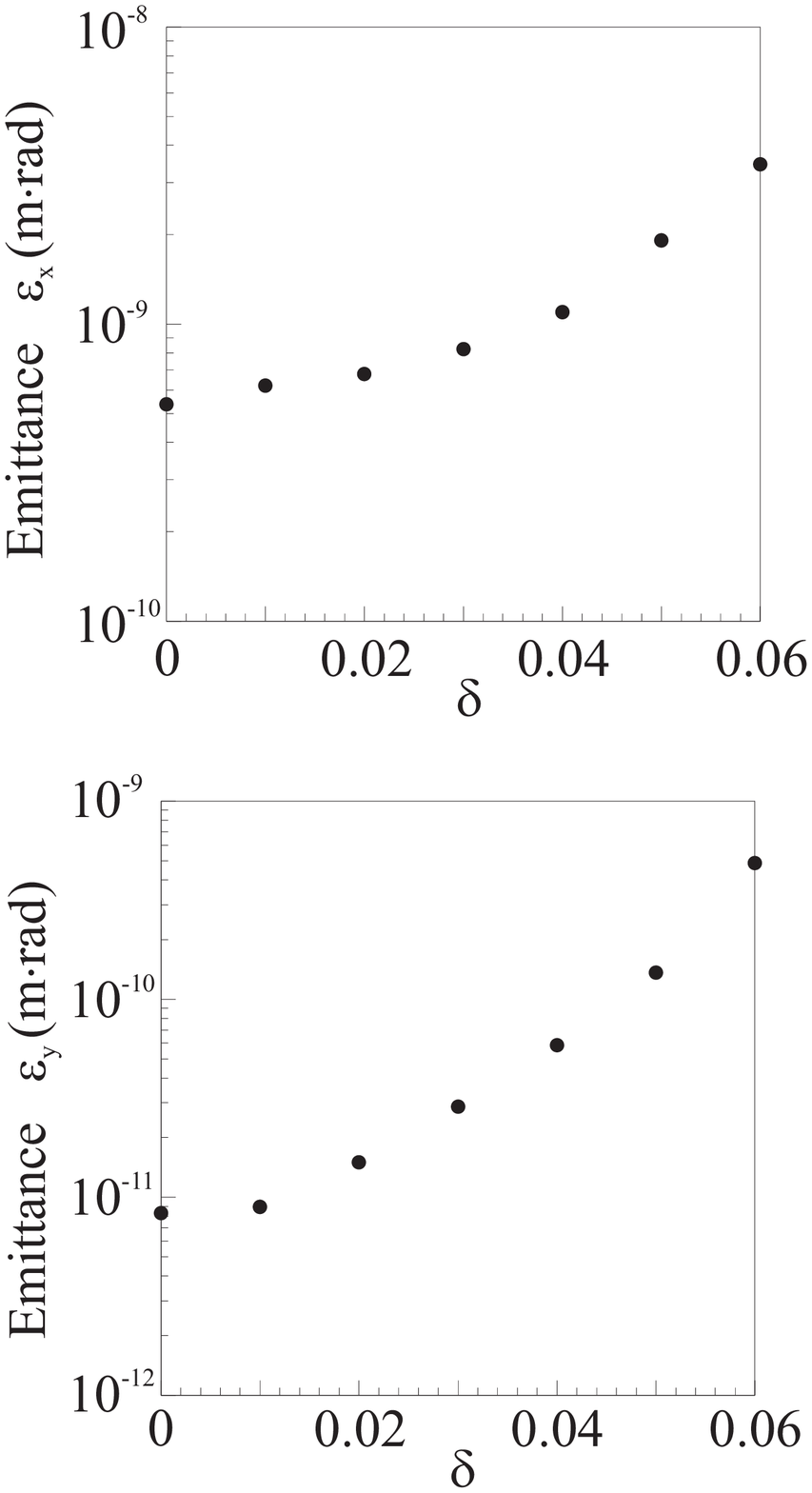,width=4.5cm}}
\vspace*{-3pt}
\fcaption{Momentum dependence of the \\
emittances in the defocus optics with \\
chromaticity correction.}
\label{fig:mdn}
\end{minipage}
\vspace*{-10pt}
\end{figure}

  The optics without chromaticity correction for the optics 2 does not
work as we saw in the previous subsection. In this subsection we apply
the chromaticity correction for the optics 2. The lattice for cooling
is designed referring to the final focus system of the future linear
colliders by K.~Oide~\cite{oid89}. The final doublet system is the
same lattice as the optics before subsection. The method of
chromaticity correction uses one family of sextupole to correct for
vertical chromaticity and moreover we added two weak sextupoles in the
lattice to correct for horizontal chromaticity. Figure~\ref{fig:dno}
shows the defocus optics with chromaticity correction for the
laser-Compton cooling. The input file is attached to
Ref.~\citenum{ohg99}. The total length of the lattice is about 63~m.

  The momentum dependence of the emittances in the defocus optics with
chromaticity correction is shown in Fig.~\ref{fig:mdn}. The 10000
particles are tracked for the transverse and longitudinal Gaussian
distribution by the MAD code. The relative energy spread $\delta$ is
changed from 0 to 0.06 with the conservation $\kappa_2~\theta_B$,
where $\kappa_2$ and $\theta_B$ are the strength of the sextupole and
the angle of the bending magnet. The initial $\beta_x$ and $\beta_y$
after laser-Compton interaction are 20~mm and 4~mm, respectively. The
final $\beta_x$ and $\beta_y$ are assumed to be 2~m and 1~m,
respectively. The initial and final $\alpha_x(y)$ with no energy
spread $\delta=0$ are 0 in this optics. After the chromaticity
correction, the chromaticity functions $\xi_x$ and $\xi_y$ are 9.3 and
1.6, respectively. If we set a limit of 200\% for
$\Delta\epsilon_i/\epsilon_i (i=x,y)$, the permissible energy spread
$\delta_x$ and $\delta_y$ are 0.040 and 0.023 which mean the momentum
band widths $\pm8\%$ and $\pm4.6\%$, respectively. By the comparison
with the results of the optics without chromaticity correction at a
limit of $200\%$ for $\Delta\epsilon_i/\epsilon_i (i=x,y)$, the
$\epsilon_y$ of the optics with chromaticity correction is about two
times larger than that of the one before subsection, but the
$\epsilon_x$ of the optics with chromaticity correction is three times
smaller than that of the one before. The results are still not
sufficient for cooling with $E_0=5$ GeV and $C=5$. These results
emphasize the need to pursue further ideas for plasma
lens~\cite{che90}.
%--------------------------------------------------------------------
%   Figure 11: The defocus optics with chromaticity correction for
%              Laser-Compton cooling.
%--------------------------------------------------------------------
%--------------------------------------------------------------------

%--------------------------------------------------------------------

\section{Laser-Compton Cooling for JLC/NLC at \boldmath$E_0=2$ GeV}             %) A SECTION HEADING

\subsection{Optics}

  For the future linear colliders, the method of laser-Compton cooling
is effective to reduce the transverse emittances after damping
rings. Where can it be placed? There are two possibilities for
JLC/NLC~\cite{yok99} as follows:
\begin{enumerate}
  \item After the first bunch compressor (BC1) and before the
pre-linac. $E_0=2$ GeV and $\sigma_z=0.5$ mm.
  \item After the second bunch compressor (BC2) and before the main
linac. $E_0=10$ GeV and $\sigma_z=0.1$ mm.
\end{enumerate}
Case~2 needs a large energy for recovery after Compton scattering and
we consider case~1 in this study. The parameters of the electron and
laser beams for laser-Compton cooling for JLC/NLC at $E_0=2$ GeV and
$C=10$ are listed in Tables~\ref{tbl:peb} and~\ref{tbl:plb}. The
energy of laser pulse is 300 J. The simulation results of the
laser-electron interaction by the CAIN code are summarized as
follows. The energy spread of the electron beam is 11\%. The decrease
of the longitudinal polarization of the electron beam is 0.038
($P_e=1.0,P_L=1.0$). The number of the scattered photons per incoming
particle and the average of the photon energy are 200 and 8.9 MeV (rms
energy 19 MeV), respectively.

%--------------------------------------------------------------------
%   Table 4 : Parameters of Defocus Optics for JLC/NLC
%--------------------------------------------------------------------
\begin{table}[htbp]
\tcaption{Parameters of the defocus optics for laser-Compton cooling for
JLC/NLC at $E_0$=2 GeV and $C=10$.}
\centerline{\footnotesize\smalllineskip
\label{tbl:pod}
\begin{tabular}{ccccc}\\
\hline
  $l$ & Length of Q1 & Field of Q1 & Aperture & Total length \\
\hline
  5~mm & 2~cm & 1.2~Tesla & 0.5~mm & 7.4~cm \\
\hline\\
\end{tabular}}
\end{table}  
%--------------------------------------------------------------------

  The electron energy after Compton scattering in case~2 is 0.2 GeV
and the strength of the final quadrupole from Eq.~(\ref{eq:3101}) is
600 m$^{-2}$. Table~\ref{tbl:pod} lists the parameters of the
defocusing optics for laser-Compton cooling for JLC/NLC at $E_0=2$ GeV
and $C=10$. The final $\beta_x$ and $\beta_y$ are assumed to be 1~m
and 0.25~m, respectively. The chromaticity functions $\xi_x$ and
$\xi_y$ are 18 and 23, respectively. Using the MAD code, the emittance
growth in the defocus optics are
\begin{equation}
\label{eq:4101}
  \Delta\epsilon_{n,x}^{\rm defocus}=\epsilon_{n,x}-\epsilon_{n,x0}\sim1.0\epsilon_{n,x0}\sim7.6\times10^{-8}~[\rm m\cdot rad],
\end{equation}
\begin{equation}
\label{eq:4102}
  \Delta\epsilon_{n,y}^{\rm defocus}=\epsilon_{n,y}-\epsilon_{n,y0}\sim1.6\epsilon_{n,y0}\sim4.6\times10^{-8}~[\rm m\cdot rad],
\end{equation}
where the normalized emittances before and after the defocus optics
are $\epsilon_{n,i0}$ and $\epsilon_{n,i}$ ($i=x,y$),
respectively. The emittance growth in the other two focus optics are
negligible.
  
\subsection{Reacceleration Linac}

  In the reacceleration linac, there are two major sources of the
emittance increase~\cite{yok99} as follows: 
\begin{enumerate}
  \item The emittance growth due to the misalignment of the quadrupole
  magnet and the energy spread.
  \item The emittance growth due to the cavity misalignment.
\end{enumerate}

  The emittance growth due to these sources in the reacceleration
linac (L-band linac) are formulated by K.~Yokoya~\cite{yok99}
\begin{equation}
\label{eq:4201}
  \Delta\epsilon_{n,x}^{\rm linac}\sim3.4\times10^{-9}~[\rm m\cdot rad]\sim0.045\it\epsilon_{n,x\rm0},
\end{equation}
\begin{equation}
\label{eq:4202}
  \Delta\epsilon_{n,y}^{\rm linac}\sim3.4\times10^{-9}~[\rm m\cdot rad]\sim0.12\it\epsilon_{n,y\rm0}.
\end{equation}

  The final emittance growth and the final emittance with $C=10$ are
\begin{equation}
\label{eq:4203}
  \Delta\epsilon_{n,x}\sim7.9\times10^{-8}~[\rm m\cdot rad]\sim1.1\it\epsilon_{n,x\rm0}~\Rightarrow\epsilon_{\it n,x}\sim\rm0.21\it\epsilon_{\it n,x\rm0},
\end{equation}
\begin{equation}
\label{eq:4204}
  \Delta\epsilon_{n,y}\sim4.9\times10^{-8}~[\rm m\cdot rad]\sim1.7\it\epsilon_{n,y\rm0}~\Rightarrow\epsilon_{\it n,y}\sim\rm0.27\it\epsilon_{\it n,y\rm0}.
\end{equation}

  The total reduction factor of the 4D transverse emittance of the
laser-Compton cooling for JLC/NLC at $E_0=2$  GeV is about 18. The
decrease of the polarization of the electron beam is 0.038 due to the 
laser-Compton interaction.

%--------------------------------------------------------------------

\section{Summary}               %) A SECTION HEADING

  We have studied the method of laser-Compton cooling of electron
beams. The effects of the laser-Compton interaction for cooling have
been evaluated by the Monte Carlo simulation. From the simulation in
the multi-stage cooling, we presented that the low emittance beams
with $\epsilon_{6N}=1.2\times10^{-19}$(m$\cdot$rad)$^3$ can be
achieved in our beam parameters. We also examined the optics with and
without chromatic correction for cooling, but the optics are not
sufficient for cooling due to the large energy spread of the electron
beams.

  The laser-Compton cooling for JLC/NLC at $E_0=2$ GeV and $C=10$ was
considered. The total reduction factor of the 4D transverse emittance
of the laser-Compton cooling is about 18. The decrease of the
polarization of the electron beam is 0.038 due to the laser-Compton
interaction.

%--------------------------------------------------------------------

\nonumsection{Acknowledgments}

We would like to thank Y.~Nosochkov, K.~Oide, T.~Takahashi, V.~Telnov,
M.~Xie, and K.~Yokoya for useful comments and discussions.

\nonumsection{References}

%--------------------------------------------------------------------

\end{document}